\documentclass[12pt]{extarticle}
\usepackage{setspace}

\usepackage[T1]{fontenc}

\usepackage[latin1]{inputenc}
\usepackage{epsfig}
\usepackage[english,french]{babel}
\usepackage{color}
\usepackage{graphicx}

\usepackage{dcolumn}

\usepackage{moreverb}

\usepackage{amsmath,amssymb,amsfonts}

\begin{document}
\numberwithin{equation}{section}
\newcommand{\boxedeqn}[1]{%
  \[\fbox{%
      \addtolength{\linewidth}{-2\fboxsep}%
      \addtolength{\linewidth}{-2\fboxrule}%
      \begin{minipage}{\linewidth}%
      \begin{equation}#1\end{equation}%
      \end{minipage}%
    }\]%
}


\newsavebox{\fmbox}
\newenvironment{fmpage}[1]
     {\begin{lrbox}{\fmbox}\begin{minipage}{#1}}
     {\end{minipage}\end{lrbox}\fbox{\usebox{\fmbox}}}

\begin{flushleft}
\title*{{\LARGE{\textbf{Superintegrability with third order integrals of motion, cubic algebras and supersymmetric quantum mechanics I:Rational function potentials}}}}
\newline
\newline
Ian Marquette
\newline
D\'epartement de physique et Centre de recherche math\'ematique,
Universit\'e de Montr\'eal,
\newline
C.P.6128, Succursale Centre-Ville, Montr\'eal, Qu\'ebec H3C 3J7,
Canada
\newline
ian.marquette@umontreal.ca
\newline
\newline
We consider a superintegrable Hamiltonian system in a
two-dimensional space with a scalar potential that allows one
quadratic and one cubic integral of motion. We construct the most
general cubic algebra and we present specific
realizations. We use them to calculate the energy spectrum. All
classical and quantum superintegrable potentials separable in
Cartesian coordinates with a third order integral are known. The
general formalism is applied to quantum reducible and irreducible rational potentials separable in
Cartesian coordinates in E2. We also discuss these potentials from the point of view of supersymmetric and PT-symmetric quantum mechanics.
\section{Introduction}
In classical mechanics a Hamiltonian system with Hamiltonian H and integrals of motion $X_{a}$
\newline
\begin{equation}
H=\frac{1}{2}g_{ik}p_{i}p_{k}+V(\vec{x},\vec{p}),\quad X_{a}=f_{a}(\vec{x},\vec{p}),\quad a=1,..., n-1 \quad,
\end{equation}
\newline
is called completely integrable (or Liouville integrable) if it
allows n integrals of motion (including the Hamiltonian) that are
well defined functions on phase space, are in involution
$\{H,X_{a}\}_{p}=0$, $\{X_{a},X_{b}\}_{p}=0$, a,b=1,...,n-1 and
are functionally independent ($\{,\}_{p}$ is a Poisson bracket). A
system is superintegrable if it is integrable and allows further
integrals of motion $Y_{b}(\vec{x},\vec{p})$, $\{H,Y_{b}\}_{p}=0$,
b=n,n+1,...,n+k that are also well defined functions on phase
space and the integrals$\{H,X_{1},...,X_{n-1},Y_{n},...,Y_{n+k}\}$
are functionally independent. A system is maximally
superintegrable if the set contains 2n-1 functions, quasi-maximally superintegrable if it contains 2n-2 and minimally
superintegrable if it contains n+1 such integrals. The integrals
$Y_{b}$ are not required to be in evolution with
$X_{1}$,...$X_{n-1}$, nor with each other. The same definitions apply in quantum mechanics but
$\{H,X_{a},Y_{b}\}$ are well defined quantum mechanical operators,
assumed to form an algebraically independent set.
\newline 
Superintegrable systems appear in many domains of physics such quantum chemistry, condensed matter and nuclear physics. The most well known examples of (maximally) superintegrable systems are the Kepler-Coulomb [1,2] system $V(\vec{x})=\frac{\alpha}{r}$ and the harmonic oscillator $V(\vec{x})=\alpha r^{2}$ [3,4]. A systematic search for superintegrable systems in two-dimensional Euclidean space $E_{2}$ was started some time ago [5,6]. In 1935 J. Drach published two articles on two-dimensional Hamiltonian systems with third order integrals of motion and found 10 such integrable classical potentials in complex Euclidean space $E_{2}(\mathbb{C})$ [7]. A systematic study of superintegrable classical and quantum system with a third order integral is more recent [8,9]. All classical and quantum potentials with a second and a third order integral of motion that separate in cartesian coordinates in the two-dimensional Euclidean space were found in Ref 9. There are 21 quantum potentials and 8 classical potentials.
\newline
The classical potentials were studied earlier [10]. In all 8 cases of superintegrable systems, separating in Cartesian
coordinates and allowing a third order integral of motion, the
integrals of motion generate a cubic Poisson algebra. In many cases
this polynomial algebra is reducible, that is it is a consequence of
the existence of a simpler algebraic structure. We have also studied trajectories and have shown that bounded trajectories are always closed for these superintegrable potentials.
\newline
The quantum case is much richer: 21 superintegrable cases of the
considered type exist, 13 of them irreducible. In this context we call a potential, or a Hamiltonian "reducible" if the third order integral is the commutator (or Poisson commutator) of two second order integrals. The potentials are
expressed in terms of rational functions in 6 cases, elliptic
functions in 2 cases and Painlev\'e transcendents [11] $P_{I}$,$P_{II}$
and $P_{IV}$ in 5 cases. 
\newline
The three reducible cases are
\newline
\newline
$V=\frac{\omega^{2}}{2}(x^{2} +y^{2})$, \quad  $V=\frac{\omega^{2}}{2}(x^{2} +y^{2}) + \frac{b}{x^{2}} +
\frac{c}{y^{2}}$, \quad $V=\frac{\omega^{2}}{2}(4x^{2} + y^{2}) + \frac{b}{y^{2}} + cx$ .
\newline
\newline
The irreducible potentials with rational function are:
\newline
\newline
Potential 1. $V=\hbar^{2}[
\frac{x^{2}+y^{2}}{8a^{4}} +
\frac{1}{(x-a)^{2}}+\frac{1}{(x+a)^{2}}]$
\newline
Potential 2. $V=\frac{\omega^{2}}{2}(9x^{2} + y^{2})  $
\newline
Potential 3. $V=\frac{\omega^{2}}{2}(9x^{2} + y^{2})+\frac{\hbar^{2}}{y^{2}}    $
\newline
Potential 4. $V=\hbar^{2}[
\frac{9x^{2}+y^{2}}{8a^{4}} +
\frac{1}{(y-a)^{2}}+\frac{1}{(y+a)^{2}}]$
\newline
Potential 5. $V=\hbar^{2}(\frac{1}{8a^{4}}[(x^{2}+y^{2})+\frac{1}{y^{2}}+\frac{1}{(x+a)^{2}}+\frac{1}{(x-a)^{2}}
]  $
\newline
Potential 6. $V=\hbar^{2}[\frac{1}{8a^{4}}(x^{2}+y^{2})+\frac{1}{(y+a)^{2}}+\frac{1}{(y-a)^{2}}
+\frac{1}{(x+a)^{2}}+\frac{1}{(x-a)^{2}} ]   $ .
\newline
\newline
It is well know that in quantum mechanics the operators commuting with the Hamiltonian, form an o(4) algebra for the hydrogen atom [3,4] and a u(3) algebra for the harmonic oscillator. We can obtain from the algebra the energy spectrum. In many cases the algebra is no longer a Lie algebra and many examples of polynomial algebras were obtained in quantum mechanics [12,13,14,15,16,17,18,19,20,21,22]. C.Daskaloyannis studied the case of the quadratic Poisson algebras of two-dimensional classical superintegrable systems and quadratic (associative) algebras of quantum superintegrable systems [17]. He shows how the quadratic algebras provide a method to obtain the energy spectrum. He uses realizations in terms of deformed oscillator algebras [18]. Potentials with a third order integral can be investiged using these techniques. 
\newline
Supersymmetry was originally introduced in the context of grand unification theory in elementary particle physics in terms of quantum field theory involving a symmetry between bosons and fermions [23]. So far there is no experimental evidence of SUSY particles. At our energies we can distinguish bosons and fermions and this symmetry should appear as a broken symmetry. Supersymmetric quantum mechanics (SUSYQM)  was introduced by E.Witten [24] as a toy model to study supersymmetry breaking. This method is related to earlier investigation of spectral properties of Sturm-Liouville differential operators by G.Darboux [25] and T.F.Moutard [26] in the 19th century. SUSYQM is also related to the factorization method that was used by E.Schrodinger in the context of the quantum harmonic oscillator [27]. The factorization method was investigated more systematically later by L.Infeld and T.F.Hull [28]. SUSYQM is now an independent field with applications to atomic, nuclear, condensed matter, statistical physics and quantum mechanics [29]. The relation with exactly solvable potentials has been discussed [30,31] and also with superintegrable potentials and quadratic algebras [32].
\newline
This paper is organized in the following way. In Section 2 we give the general form of the cubic algebra for two-dimensional systems with a quadratic and a cubic operator that commute with the Hamiltonian. We give a realization of the cubic algebra in terms of parafermionic oscillator algebras. We study the finite dimensional representations of the cubic algebra. In Section 3 we apply this method to the case of irreducible potentials separable in Cartesian coordinates in $E_{2}$ with a third order integral. In Section 4 we investigate the irreducible potentials from the point of view of supersymmetric quantum mechanics. In Section 5 we give the generating spectrum algebra of the irreducible Potential 1. In Section 6 we investigate the complexification of the irreducible Potential 1. (All other cases can we obtained from Potential 1).
\newline
\section{Cubic and parafermionic algebras }
We consider a quantum superintegrable system with a
quadratic Hamiltonian and one second order and one third order
integral of motion
\begin{equation}
H = a(q_{1},q_{2})P_{1}^{2} + 2b(q_{1},q_{2})P_{1}P_{2} +
c(q_{1},q_{2})P_{2}^{2} + V(q_{1},q_{2})
\end{equation}
\begin{equation*}
A = d(q_{1},q_{2})P_{1}^{2} +2e(q_{1},q_{2})P_{1}P_{2} +f(q_{1},q_{2})P_{2}^{2} + g(q_{1},q_{2})P_{1} + h(q_{1},q_{2})P_{2}
+ Q(q_{1},q_{2})
\end{equation*}
\begin{equation*}
B =  u(q_{1},q_{2})P_{1}^{3} +
3v(q_{1},q_{2})P_{1}^{2}P_{2} + 3w(q_{1},q_{2})P_{1}P_{2}^{2}
+x(q_{1},q_{2})P_{2}^{3}+j(q_{1},q_{2})P_{1}^{2}
\end{equation*}
\begin{equation*}
+2k(q_{1},q_{2})P_{1}P_{2} + l(q_{1},q_{2})P_{2}^{2}+m(q_{1},q_{2})P_{1} + n(q_{1},q_{2})P_{2} + S(q_{1},q_{2})\quad ,
\end{equation*}
with
\newline
\begin{equation}
P_{1}=-i\hbar \partial_{1} , P_{2}=-i\hbar \partial_{2}\quad ,
\end{equation}
\begin{equation}
[H,A]=[H,B]=0 \quad .
\end{equation}
\newline
We assume that our integrals close in a cubic algebra. This is the quantum version of the cubic Poisson algebra obtained earlier [10] and the cubic generalization of the quadratic algebra studied by C.Daskaloyannis [17]. The most general form of such an algebra is
\newline
\[ [A,B]=C \]
\begin{equation}
[A,C] = \alpha A^{2} + \beta \{A,B\} + \gamma A + \delta B +
\epsilon
\end{equation}
\[ [B,C] = \mu A^{3} + \nu A^{2} + \rho B^{2} + \sigma \{A,B\} + \xi A +
\eta B + \zeta \]
\newline
where $\{,\}$ denotes an anticommutator. The coefficients $ \alpha$ , $\beta$ and $\mu$ are constants, but
the other ones can be polynomials in the Hamiltonian H. The degrees
of these polynomials are dictated by the fact that H and A are
second order polynomials in the momenta and B is a third order one.
Hence C can be a fourth order polynomial. The Jacobi identity $[A,[B,C]]=[B,[A,C]] $ implies $ \rho = -\beta$
, $\sigma= - \alpha$ and $\eta = - \gamma $. We obtain
\newline
\begin{subequations}
\begin{equation}
[A,B]=C 
\end{equation}
\begin{equation}
[A,C]=\alpha A^{2} + \beta \{A,B\} + \gamma A + \delta B + \epsilon
\end{equation}
\begin{equation}
[B,C]=\mu A^{3} + \nu A^{2} - \beta B^{2} - \alpha \{A,B\} + \xi A -
\gamma B + \zeta \quad . 
\end{equation}
\end{subequations}
\newline
For the polynomials on the left and right sides of Eq.(2.4) and Eq.(2.5) to have the same degree we must have
\newline
\begin{equation}
\alpha=\alpha_{0},\quad \beta=\beta_{0},\quad \mu=\mu_{0}
\end{equation}
\[\gamma=\gamma_{0}+\gamma_{1}H,\quad \delta=\delta_{0}+\delta_{1}H,\quad \epsilon=\epsilon_{0}+\epsilon_{1}H+\epsilon_{2}H^{2}\]
\[\nu=\nu_{0}+\nu_{1}H,\quad \xi=\xi_{0}+\xi_{1}H+\xi_{2}H^{2}\]
\[\zeta=\zeta_{0}+\zeta_{1}H+\zeta_{2}H^{2}+\zeta_{3}H^{3},\]
\newline
where $\alpha_{0}$, ... , $\zeta_{3}$ are constants.
\newline
The Casimir operator of a polynomial algebra is an operator that
commutes with all elements of the algebra:
\begin{equation}
[K,A]=[K,B]=[K,C]=0
\end{equation}
and this implies
\newline
\[ K = C^{2} - \alpha \{A^{2},B\} - \beta \{A,B^{2}\} + (\alpha
\beta - \gamma)\{A,B\} + (\beta^{2} - \delta)B^{2} \]
\begin{equation}
(+\beta \gamma - 2\epsilon)B+\frac{\mu}{2}A^{4} +
\frac{2}{3}(\nu+\mu \beta)A^{3}+(-\frac{1}{6}\mu \beta^{2} +
\frac{\beta \nu}{3} + \frac{\delta \mu}{2} + \alpha^{2} + \xi)A^{2}
\end{equation}
\[+(-\frac{1}{6}\mu \beta \delta + \frac{\delta \nu}{3} + \alpha
\gamma + 2\zeta)A \quad . \]
\newline
Ultimatly, the Casimir operator will be a function of the Hamiltonian alone. We construct a realization of the cubic algebra in terms of a deformed oscillator algebra [17,18] $ \{b^{t},b,N\}$ which satisfies the relation
\newline
\begin{equation}
[N,b^{t}]=b^{t} ,\quad [N,b]=-b ,\quad b^{t}b=\Phi(N) ,\quad
bb^{t}=\Phi(N+1) \quad .
\end{equation}
\newline
$\Phi(N)$ is called the "structure function". Following C.Daskaloyannis [17] we request $\Phi(N)$ to be a real function and impose $ \Phi(0)=0$ and $\Phi(N) > 0$ for $N > 0 $. We construct a Fock type representation for the deformed oscillator algebra with a Fock basis $|n>$ ,
n=0,1,2... satisfying
\newline
\begin{equation}
N|n>=n|n>,\quad b^{t}|n>=\sqrt{\Phi(N+1)}|n+1>,
\end{equation}
\begin{equation}
b|0>=0,\quad b|n>=\sqrt{\Phi(N)}|n-1> \quad .
\end{equation}
\newline
To obtain a finite-dimensional representation we request $\Phi(p+1)=0$.
\newline
\newline
Let us show that there exists a realization of the form:
\newline
\begin{equation}
A=A(N),\quad B=b(N)+b^{t}\rho(N)+\rho(N)b \quad .
\end{equation}
\newline
The functions A(N) , b(N) et $\rho(N)$ will be determined by the
cubic algebra. We have by Eq.(2.5.a)
\begin{equation}
C=[A,B]=b^{t}\bigtriangleup A(N) \rho(N) - \rho(N)\bigtriangleup A(N)b,
\end{equation}
\newline
where $\bigtriangleup A(N)$ is define to be $\bigtriangleup A(N)= A(N+1) - A(N)$. When we insert Eq.(2.12) into Eq.(2.5b)
we obtain two equations that allow us to determine A(N) and b(N)
\newline
\begin{equation}
\bigtriangleup A(N)^{2} = \gamma(A(N+1)+A(N))+\epsilon
\end{equation}
\[\alpha A(N)^{2} + 2\beta A(N)b(N) + \gamma A(N) + \delta
b(N) + \epsilon = 0\quad .\]
\newline
Two distinct possibilities occur.
\newline
\textbf{Case 1}: $ \beta \neq 0$. We find the following solution
\begin{equation}
A(N) = \frac{\beta}{2}((N+u)^{2}- \frac{1}{4} -
\frac{\delta}{\beta^{2}})
\end{equation}
\[ b(N) = \frac{\alpha}{4}((N+u)^{2}-\frac{1}{4})+ \frac{\alpha
\delta - \gamma \beta}{2\beta^{2}}-\frac{\alpha \delta^{2}-2\gamma \delta \beta +
4\beta^{2}\epsilon}{4\beta^{4}}\frac{1}{(N+u)^{2}-\frac{1}{4}}\quad . \]
\newline
\newline
The constant u will be determined below using the fact that we
require that the deformed oscillator algebras should be nilpotent. Eq.(2.5c) gives us
\newline
\begin{equation}
2\Phi(N+1)( \bigtriangleup A(N) + \frac{\gamma}{2})\rho(N) -
2\Phi(N)(\bigtriangleup A(N-1) - \frac{\gamma}{2})\rho(N-1)
\end{equation}
\[ =  \mu A(N)^3 + \nu A(N)^2 - \beta b(N)^2 - 2\alpha A(N)b(N) + \xi A(N) -
\gamma b(N) + \zeta  \]
\newline
and the Casimir operator is now realized as
\newline
\begin{equation}
K = \Phi(N+1)(\beta^{2}-\delta-2\beta A(N)-\bigtriangleup
A(N)^{2})\rho(N)
\end{equation}
\[+\Phi(N)(\beta^{2}-\delta-2\beta A(N) -
\bigtriangleup A(N-1)^{2})\rho(N-1)-2\alpha A(N)^{2}b(N) \]
\[+(\beta^{2}-\delta-2\beta A(N))b(N)^{2} + 2(\alpha \beta
-\gamma)A(N)b(N) + (\beta \gamma - 2\epsilon)b(N) +
\frac{\mu}{2}A(N)^{4}\]
\[+\frac{2}{3}(\nu+\mu \beta)A(N)^{3}+(-\frac{1}{6}\mu \beta^{2} +
\frac{\beta \nu}{3} + \frac{\delta \mu}{2} + \alpha^{2} +
\epsilon)A(N)^{2}+(-\frac{1}{6}\mu \beta \delta + \frac{\delta a}{3} + \alpha
\gamma + 2\zeta)A(N) \quad . \]
\newline
Finally the structure function is
\newline
\[\Phi(N)=\frac{1}{\rho(N-1)(\bigtriangleup
A(N-1)-\frac{\beta}{2})(f)+(\bigtriangleup
A(N)+\frac{\beta}{2})(g})\]
\[ [(\bigtriangleup A(N)+\frac{\beta}{2})(K+2\alpha
A(N)^{2}b(N)-(\beta^{2}-\delta-2\beta A(N))b(N)^{2}\]
\begin{equation}
-2(\alpha \beta-\gamma)A(N)b(N)-(\beta \gamma
-2\epsilon)b(N)-\frac{\mu}{2}A(N)^{4}-\frac{2}{3}(\nu+\mu
\beta)A(N)^{3}
\end{equation}
\[-(-\frac{1}{6}\mu \beta^{2}+\frac{\beta
\nu}{3}+\alpha^{2}+\xi)A(N)^{2}-(-\frac{1}{6}\mu \beta \delta
+\frac{\delta \nu}{3} + \alpha \gamma + 2\zeta)A(N))\]
\[-\frac{1}{2}(\beta^{2}-\delta-2\beta A(N)-\bigtriangleup
A(N)^{2})(\mu A(N)^{3}+\nu A(N)^{2}-\beta b(N)^{2}-2\alpha
A(N)b(N)+\xi A(N)-\gamma b(N)+\zeta)] \quad ,\]
with
\begin{equation}
f=\beta^{2}-\delta-2\beta
A(N)-\bigtriangleup A(N)^{2},\quad g=\beta^{2}-\delta-2\beta A(N)-\bigtriangleup
A(N-1)^{2}\quad .
\end{equation}
\newline
Thus the structure function depends only on the function $\rho$.
This function can be arbitrarily chosen and does not influence the
spectrum. We choose $\rho$ to obtain a structure function that is a polynomial in N, namely we put
\newline
\begin{equation}
\rho(N) = \frac{1}{3*2^{12}\beta^{8}(N+u)(1+N+u)(1+2(N+u))^{2}} .
\end{equation}
From our expressions for A(N) , b(N) and $\rho(N)$, the third
relation of the cubic associative algebra and the expression of the
Casimir operator we find the structure function $\Phi(N)$. For the
Case 1 the structure function is a polynomial of order 10 in N. The
coefficients of this polynomial are functions of $\alpha$, $\beta$,
$\mu$, $\gamma$, $\delta$, $\epsilon$, $\nu$, $\xi$ and $\zeta$. We give the formula in the Appendix.
\newline
\newline
\textbf{Case 2}: For $\beta=0$ and $\delta \neq 0$ we get the solution
\begin{equation}
A(N)= \sqrt{\delta}(N+u), b(N)= -\alpha(N+u)^{2} -
\frac{\gamma}{\sqrt{\delta}}(N+u) - \frac{\epsilon}{\delta}\quad .
\end{equation}
\newline
We choose a trivial expression $\rho(N)=1$. The explicit
expression of the structure function for this case is
\newline
\begin{equation}
\Phi(N) = (\frac{K}{-4\delta}-\frac{\gamma
\epsilon}{4\delta^{\frac{3}{2}}}-\frac{\zeta}{4\sqrt{\delta}} +
\frac{\epsilon^{2}}{4\delta^{2}})
\end{equation}
\[+(\frac{-\alpha
\epsilon}{2\delta}-\frac{\xi}{4}-\frac{\gamma^{2}}{4\delta}+\frac{\gamma
\epsilon}{2 \delta^{\frac{3}{2}}}+\frac{\alpha
\gamma}{4\sqrt{\delta}}+\frac{\zeta}{2\sqrt{\delta}}+\frac{\nu\sqrt{\delta}}{12})(N+u)\]
\[+(\frac{-\nu\sqrt{\delta}}{4}-\frac{3\alpha
\gamma}{4\sqrt{\delta}}+\frac{\gamma^{2}}{4\delta}+\frac{\epsilon
\alpha}{2\delta}+\frac{\alpha^{2}}{4}+\frac{\xi}{4}+\frac{\mu\delta}{8})(N+u)^{2}\]
\[+(\frac{-\alpha^{2}}{2}+\frac{\gamma
\alpha}{2\delta^{\frac{1}{2}}}+\frac{\nu\sqrt{\delta}}{6}-\frac{\mu
\delta}{4})(N+u)^{3}+(\frac{\alpha^{2}}{4}+\frac{\mu\delta}{8})(N+u)^{4} \quad . \]
\newline
We use a parafermionic realization in which the parafermionic number
operator N and the Casimir operator K are diagonal. The basis of
this representation is the Fock basis for the parafermionic
oscillator. The vector $|k,n>,n=0,1,2...$ satisfies the following
relations:
\newline
\begin{equation}
N|k,n>=n|k,n>, K|k,n>=k|k,n> \quad .
\end{equation}
The vectors $|k,n>$ are also eigenvectors of the generator A.
\[ A|k,n>=A(k,n)|k,n>, \]
\begin{equation}
A(k,n)=\frac{\beta}{2}((n+u)^{2}-\frac{1}{4}-\frac{\delta}{\beta^{2}})
,\quad \beta \neq 0 ,
\end{equation}
\[ A(k,n) = \sqrt{\delta}(n+u) ,\quad \beta = 0,\quad \delta \neq 0 \quad . \]
\newline
We have the following constraints for the structure function
\begin{equation}
\Phi(0,u,k)=0 ,\quad \Phi(p+1,u,k)=0 \quad .
\end{equation}
\newline
With these two relations we can find the energy spectrum. Many
solutions for the system exist. Unitary representations of the
deformed parafermionic oscillator obey the constraint
$\Phi(x)
>0 $ for x=1,2,...,p .
\newline
\newline
There are other conditions that should be imposed. The representations should be constrained by the differential character of the Hamiltonian and the integrals. For example the mean energy should be greater than the minimum of the potential
\newline
\begin{equation}
<H> \geq min V
\end{equation}
This restriction and possibly other ones coming from the differential operator character of the integrals should be taken into consideration to exclude spurious states.
\section{Irreducible rational function potentials}
In the case of the three reducible superintegrable potentials the cubic algebra is a direct consequence of a simpler algebraic structure. The first potential $V=\frac{\omega^{2}}{2}(x^{2} +y^{2})$ is the well known isotropic harmonic oscillator. We can construct the quadratic or the cubic algebra from the Lie algebra as in the classical case [10]. The eigenfunctions of the harmonic oscillator are well known and are given in terms of the Hermite polynomials. The two other reducible potentials $V=\frac{\omega^{2}}{2}(x^{2} +y^{2}) + \frac{b}{x^{2}} + \frac{c}{y^{2}}$ and $V=\frac{\omega^{2}}{2}(4x^{2} + y^{2}) + \frac{b}{y^{2}} + cx$ are two of the four types of potentials found a long time ago [6]. There is no Lie algebra in these cases but a quadratic algebra [17] and we can obtain the cubic algebra directly from this algebra. We obtain from the cubic algebra the same unitary representations that were obtained from the quadratic algebra [17]. 
\newline
In this section we will apply to the irreducible quantum potentials resuts of the Section 2 and give all unitary representations and the corresponding energy spectra. Notice that in all case we have $\beta=0$ so only Case 2 of Section 2 occurs.
\newline
\newline
\textbf{Potential 1}: $V=\hbar^{2}(\frac{x^{2}+y^{2}}{8a^{4}} + \frac{1}{(x-a)^{2}}+\frac{1}{(x+a)^{2}})$
\newline
\newline
This potential has the following two integrals 
\begin{equation}
A = P_{x}^{2} - P_{y}^{2} + 2\hbar^{2}( \frac{x^{2}-y^{2}}{8a^{4}} +
\frac{1}{(x-a)^{2}}+\frac{1}{(x+a)^{2}})
\end{equation}
\begin{equation} B = \frac{1}{2}\{L,P_{x}^{2}\} + \frac{1}{2}\hbar^{2}\{y(
\frac{4a^{2}-x^{2}}{4a^{4}} -
\frac{6(x^{2}+a^{2})}{(x^{2}-a^{2})^{2})}),P_{x}\}
\end{equation}
\begin{equation*}
+ \frac{1}{2}\hbar^{2}\{x(\frac{(x^{2}-4a^{2})}{4a^{4}} -
\frac{2}{x^{2}-a^{2}} + \frac{4(x^{2}+a^{2})}{(x^{2}-a^{2})^{2}}
),P_{y}\}.
\end{equation*}
\newline
The integrals A,B and H give rise to the following cubic algebra and Casimir operator
\newline
\begin{equation}
[A,B]=C , \quad [A,C]=\frac{4h^{4}}{a^{4}}B
\end{equation}
\[ [B,C]= -2\hbar^{2}A^{3} - 6\hbar^{2}A^{2}H + 8\hbar^{2}H^{3}
+ 6\frac{\hbar^{4}}{a^{2}}A^{2} + 8\frac{\hbar^{4}}{a^{2}}HA -
8\frac{\hbar^{4}}{a^{2}}H^{2} + 2\frac{\hbar^{6}}{a^{4}}A - 2\frac{\hbar^{6}}{a^{4}}H -
6\frac{\hbar^{8}}{a^{6}} \]
\begin{equation}
K = -16\hbar^{2}H^{4} + 32\frac{\hbar^{4}}{a^{2}}H^{3} +
16\frac{\hbar^{6}}{a^{4}}H^{2} - 40\frac{\hbar^{8}}{a^{6}}H -
3\frac{\hbar^{10}}{a^{8}}\quad .
\end{equation}
\newline
The structure function is given by the expression 
\newline
\begin{equation}
\Phi(x)=(\frac{-\hbar^{8}}{a^{4}})(x+u -
(\frac{-a^{2}E}{\hbar^{2}}-\frac{1}{2}))(x+u -
(\frac{a^{2}E}{\hbar^{2}}+\frac{1}{2}))(x+u -
(\frac{-a^{2}E}{\hbar^{2}}+\frac{3}{2}))(x+u -
(\frac{-a^{2}E}{\hbar^{2}}+\frac{5}{2}))\quad .
\end{equation}
\newline
There are three unitary representations. The first unitary solution is for $a=ia_{0},\quad a_{0}\in \mathbb{R}$. From the condition $\Phi(0,u,k)=0$ we find $u=\frac{-a_{0}^{2}E}{\hbar^{2}}+\frac{1}{2}$. The second constraint $\Phi(p+1,u,k)=0$ implies
\newline
\begin{equation}
E=\frac{\hbar^{2}(p+2)}{2a_{0}^{2}},\quad
\Phi(x)=(\frac{\hbar^{8}}{a_{0}^{4}})x(p+1-x)(p+3-x)(p+4-x),
\end{equation}
where $p\in \mathbb{N}$. We have $\Phi(p+1)=0$ which means that the unitary representations have dimension p+1. This is also the degeneracy of the energy levels.
\newline
The second unitary solution for $a=ia_{0}, a_{0} \in \mathbb{R}$. We have $u=\frac{a_{0}^{2}E}{\hbar^{2}}-\frac{1}{2}$ and 
\newline
\begin{equation}
E=-\frac{\hbar^{2}(p)}{2a_{0}^{2}},\quad
\Phi(x)=(\frac{\hbar^{8}}{a_{0}^{4}})x(p+1-x)(3-x)(2-x)\quad,
\end{equation}
\newline
valid only for p=0,1.
\newline
We have 
\begin{equation}
E \geq min V=V(0,0)=\frac{-2\hbar^{2}}{a_{0}^{2}}\quad ,
\end{equation}
so this is a physically meaningful solution. A third unitary solution exists this time for $a \in \mathbb{R}$. We have $u=\frac{-a^{2}E}{\hbar^{2}}+\frac{5}{2}$ and
\begin{equation}
E=\frac{\hbar^{2}(p+3)}{2a^{2}},\quad
\Phi(x)=(\frac{\hbar^{8}}{a^{4}})x(p+1-x)(x+1)(x+3)\quad .
\end{equation}
\newline
\newline
\textbf{Potential 2}: $V=\frac{\omega^{2}}{2}(9x^{2} + y^{2})$
\newline
\newline
This potential has the two integrals
\begin{equation}
A =P_{x}^{2} - P_{y}^{2} + \omega^{2}(9x^{2} - y^{2})
\end{equation}
\[ B=\frac{1}{2}\{L,P_{y}^{2}\} + \frac{\omega^{2}}{6}\{y^{3},P_{x}\} -
\frac{3\omega^{2}}{2}\{xy^{2},P_{y}\}\quad . \]
\newline
The cubic algebra and Casimir operator of this system are
\newline
\begin{equation}
[A,B]=C, \quad [A,C]=144\omega^{2}\hbar^{2} B
\end{equation}
\[ [B,C]= -2\hbar^{2}A^{3} + 6\hbar^{2}HA^{2} - 8\hbar^{2}H^{3} -
56\omega^{2}\hbar^{4}A + 72\omega^{2}\hbar^{4}H  \]
\begin{equation}
K = -16\hbar^{2}H^{4} + 64\omega^{2}\hbar^{4}H^{2} +
720\omega^{4}\hbar^{6} \quad .
\end{equation}
\newline
The structure function is
\newline
\begin{equation}
\end{equation}
\[ \Phi(x) = (-36\omega^{2}h^{4})( x+u -
(\frac{-E}{6\omega\hbar}+\frac{1}{2}))( x+u -
(\frac{E}{6\omega\hbar}+\frac{1}{6}))( x+u -
(\frac{E}{6\omega\hbar}+\frac{1}{2}))\]
\[( x+u - (\frac{E}{6\omega\hbar}+\frac{5}{6}))\quad . \]
We use the two constraints given by the Eq.(2.25). We obtain $u = \frac{-E}{6\omega\hbar}+\frac{1}{2}$ and three unitary representations with the corresponding energy spectra
\newline
\begin{equation}
E = 3\omega\hbar (p+\frac{2}{3}),\quad \Phi(x)=(36\omega^{2}h^{4})x(p+1-x)(p+\frac{2}{3}-x)(p+\frac{1}{3}-x)
\end{equation}
\begin{equation}
E = 3\omega\hbar (p+1),\quad \Phi(x)=(36\omega^{2}h^{4})x(p+1-x)(p+\frac{2}{3}-x)(p+\frac{4}{3}-x)
\end{equation}
\begin{equation}
E = 3\omega\hbar (p+\frac{4}{3}),\quad \Phi(x)=(36\omega^{2}h^{4})x(p+1-x)(p+\frac{5}{3}-x)(p+\frac{4}{3}-x).
\end{equation}
\newline
These results coincide with those obtained by solving the Schrödinger equation using separation of variable. The eigenfunctions are well known and given by
\begin{equation}
\phi_{k_{1}}(x)=\frac{1}{\sqrt{2^{k_{1}}k_{1}!}}(\frac{3\omega}{\pi\hbar})^{\frac{1}{4}}e^{\frac{-3\omega}{2\hbar}x^{2}}H_{k_{1}}(\sqrt{\frac{3\omega}{\hbar}}x) ,
\end{equation}
\begin{equation}
\phi_{k_{2}}(y)=\frac{1}{\sqrt{2^{k_{2}}k_{2}!}}(\frac{\omega}{\pi\hbar})^{\frac{1}{4}}e^{\frac{-\omega}{2\hbar}y^{2}}H_{k_{2}}(\sqrt{\frac{\omega}{\hbar}}y) \quad ,
\end{equation}
where $H_{k}$ are Hermite polynomials. The corresponding energy spectrum is
\begin{equation}
E=\omega\hbar (3k_{1}+k_{2}+2)
\end{equation}
\newline
\textbf{Potential 3}: $V=\frac{\omega^{2}}{2}(9x^{2} + y^{2})+\frac{\hbar^{2}}{y^{2}}$
\newline
\newline
The two integrals of this potential are 
\begin{equation}
A=P_{x}^{2} - P_{y}^{2} + \omega^{2}(9x^{2} -
y^{2})-\frac{2\hbar^{2}}{y^{2}}
\end{equation}
\[ B = \frac{1}{2}\{L,P_{y}^{2}\} + \frac{1}{2}\{\frac{\omega^{2}y^{3}}{3}-\frac{\hbar^{2}}{y},P_{x}\}
+ \frac{1}{2}\{3x(-\omega^{2}y^{2}+\frac{\hbar^{2}}{y^{2}},P_{y}\}. \]
\newline
The cubic algebra and the Casimir operator are
\newline
\begin{equation}
[A,B]=C \quad [A,C]=144\omega^{2}\hbar^{2} B
\end{equation}
\[ [B,C]= -2\hbar^{2}A^{3} + 6\hbar^{2}HA^{2} - 8\hbar^{2}H^{3} -
8\omega^{2}\hbar^{4}A + 72\omega^{2}\hbar^{4}H  \]
\begin{equation}
K = -16\hbar^{2}H^{4} + 256\omega^{2}\hbar^{4}H^{2} -
1008\omega^{4}\hbar^{6} \quad .
\end{equation}
\newline
The structure function is
\newline
\begin{equation}
\end{equation}
\[ \Phi(x) = (-36\omega^{2}h^{4})( x+u -
(\frac{-E}{6\omega\hbar}+\frac{1}{2}))( x+u -
(\frac{E}{6\omega\hbar}-\frac{1}{6}))( x+u -
(\frac{E}{6\omega\hbar}+\frac{1}{2}))\]
\[( x+u - (\frac{E}{6\omega\hbar}+\frac{7}{6})) \quad . \]
\newline
Using Eq.(2.25) we obtain $u=\frac{-E}{6\omega\hbar}+\frac{1}{2}$ and two unitary representations
\newline
\begin{equation}
\Phi(x) =
(36\omega^{2}\hbar^{4})x(p+\frac{5}{3}-x)(p+1-x)(p+\frac{7}{3}-x), \quad 
E=3\omega\hbar(p+\frac{5}{3}),
\end{equation}
\begin{equation}
\Phi(x) =
(36\omega^{2}\hbar^{4})x(p+\frac{1}{3}-x)(p+\frac{5}{3}-x)(p+1-x), \quad
E=3\omega\hbar(p+1) \quad .
\end{equation}
\newline
These results are corroborated by those obtained when we use separation of variable and solve the Schrödinger equation. The eigenfunctions are well know are given by 
\begin{equation}
\phi_{k_{1}}(x)=\frac{1}{\sqrt{2^{k_{1}}k_{1}!}}(\frac{3\omega}{\pi\hbar})^{\frac{1}{4}}e^{\frac{-3\omega}{2\hbar}x^{2}}H_{k_{1}}(\sqrt{\frac{3\omega}{\hbar}}x),
\end{equation}
\begin{equation}
\phi_{k_{2}}(y)=(\frac{\omega}{\hbar})(\frac{k_{2}!((\frac{\omega}{\hbar})^{\frac{3}{2}})}{\Gamma(k_{2}+\frac{5}{2})})^{\frac{1}{2}}e^{-\frac{\omega}{2\hbar}y^{2}}y^{2}L_{k_{2}}^{\frac{3}{2}}(\frac{\omega}{\hbar}y^{2}).
\end{equation}
where $L_{n}^{\alpha}$ is a Laguerre polynomial. The corresponding energy spectrum is
\begin{equation}
E=\omega\hbar(3k_{1}+2k_{2}+4)
\end{equation}
\newline
\textbf{Potential 4}: $V=\hbar^{2}(\frac{9x^{2}+y^{2}}{8a^{4}} + \frac{1}{(y-a)^{2}}+\frac{1}{(y+a)^{2}})$
\newline
\newline
The two integrals are given by the formulas
\begin{equation}
A =P_{x}^{2} - P_{y}^{2} + 2\hbar^{2}( \frac{9x^{2}-y^{2}}{8a^{4}} +
\frac{1}{(y-a)^{2}}+\frac{1}{(y+a)^{2}})
\end{equation}
\begin{equation}
B = \frac{1}{2}\{L,P_{y}^{2}\} + \frac{1}{2}\hbar^{2}\{y(
\frac{y^{2}}{12a^{4}} - \frac{8a^{2}}{(y^{2}-a^{2})^{2}} -
\frac{2}{y^{2}-a^{2}} ),P_{y}\}
\end{equation}
\[ + \frac{1}{2}\hbar^{2}\{x(\frac{8(y^{2}+a^{2})}{(y^{2}-a^{2})^{2}}-
\frac{y^{2}}{a^{4}}  ),P_{y}\}\]
\newline
The cubic algebra and the Casimir operator are 
\newline
\begin{equation}
[A,B]=C, \quad [A,C]=\frac{36h^{4}}{a^{4}}B
\end{equation}
\[ [B,C]= -2\hbar^{2}A^{3} - 6\hbar^{2}A^{2}H - 8\hbar^{2}H^{3}+ 10\frac{\hbar^{6}}{a^{4}}A + 18\frac{\hbar^{6}}{a^{4}}H -
24\frac{\hbar^{8}}{a^{6}} \]
\begin{equation}
K = -16\hbar^{2}H^{4}  + 112\frac{\hbar^{6}}{a^{4}}H^{2} +
96\frac{\hbar^{8}}{a^{6}}H - 171\frac{\hbar^{10}}{a^{8}}\quad .
\end{equation}
\newline
The structure function is
\newline
\begin{equation}
\Phi(x)=\frac{-9\hbar^{6}}{a^{4}}(x+u
-(\frac{a^{2}E}{3\hbar^{2}}-\frac{1}{2}
))(x+u-(\frac{-a^{2}E}{3\hbar^{2}}+\frac{1}{2}
))(x+u-(\frac{a^{2}E}{3\hbar^{2}}+\frac{5}{6}
))(x+u-(\frac{a^{2}E}{3\hbar^{2}}+\frac{7}{6}  ))\quad .
\end{equation}
\newline
For the case $ a=ia_{0},\quad a_{0}\in \mathbb{R}$  we get the three following unitary representations
\newline
\begin{equation}
\Phi(x)=\frac{9\hbar^{6}}{a_{0}^{4}}x(p+1-x)(x-\frac{4}{3})(x-\frac{5}{3}),\quad
E =\frac{3\hbar^{2}(p)}{2a_{0}^{2}}
\end{equation}
\begin{equation}
\Phi(x)=\frac{36\hbar^{6}}{a_{0}^{4}}x(p+1-x)(x+\frac{4}{3})(x-\frac{1}{3}),\quad
E =\frac{3\hbar^{2}(p+\frac{4}{3})}{2a_{0}^{2}}
\end{equation}
\begin{equation}
\Phi(x)=\frac{36\hbar^{6}}{a_{0}^{4}}x(p+1-x)(x+\frac{2}{3})(x+\frac{1}{3}),\quad
E =\frac{3\hbar^{2}(p+\frac{5}{3})}{2a_{0}^{2}}\quad .
\end{equation}
\newline
For the case $a \in \mathbb{R}$ we get the three unitary representations
\newline
\begin{equation}
\Phi(x)=\frac{9\hbar^{6}}{a^{4}}x(p+1-x)(p+\frac{7}{3}-x)(p+\frac{8}{3}-x),\quad
E =\frac{3\hbar^{2}(p+2)}{2a^{2}}
\end{equation}
\begin{equation}
\Phi(x)=\frac{9\hbar^{6}}{a^{4}}x(p+1-x)(p+\frac{4}{3}-x)(p-\frac{1}{3}-x),\quad
E =\frac{3\hbar^{2}(p+\frac{2}{3})}{2a^{2}}
\end{equation}
\begin{equation}
\Phi(x)=\frac{9\hbar^{6}}{a^{4}}x(p+1-x)(p+\frac{2}{3}-x)(p-\frac{2}{3}-x),\quad
E =\frac{3\hbar^{2}(p+\frac{1}{3})}{2a^{2}} \quad .
\end{equation}
\newline
\newline
The Potential 5 $V=\hbar^{2}(\frac{1}{8a^{4}}[(x^{2}+y^{2})+\frac{1}{y^{2}}+\frac{1}{(x+a)^{2}}+\frac{1}{(x-a)^{2}}
]  $ and Potential 6 $V=\hbar^{2}[\frac{1}{8a^{4}}(x^{2}+y^{2})+\frac{1}{(y+a)^{2}}+\frac{1}{(y-a)^{2}}
+\frac{1}{(x+a)^{2}}+\frac{1}{(x-a)^{2}} ]   $ are particular. Their integrals of motion A,B and C do not close in a finite cubic algebra. Closure at a higher order remains to be investigated. In these cases, we have the separation of variables and the unidimensional parts are related to the Potential 1 and Potential 4 and their spectra. We will see also in the next section that we can obtain information using supersymmetric quantum mechanics.
\newline
\section{Supersymmetric quantum mechanics}
In this section we will investigate a relation between SUSYQM [24] and superintegrable systems with a third order integral of motion . Let us recall some aspects of supersymmetric quantum mechanics.  We define two first order operators
\newline
\begin{equation}
A=\frac{\hbar}{\sqrt{2}}\frac{d}{dx}+W(x),\quad A^{\dagger}=-\frac{\hbar}{\sqrt{2}}\frac{d}{dx}+W(x) \quad .
\end{equation}
We consider the following two Hamiltonians which are called "superpartners"
\begin{equation}
H_{1}=A^{\dagger}A=-\frac{\hbar^{2}}{2}\frac{d^{2}}{dx^{2}}+W^{2}-\frac{\hbar}{\sqrt{2}}W',\quad H_{2}= AA^{\dagger}=-\frac{\hbar^{2}}{2}\frac{d^{2}}{dx^{2}}+W^{2}+\frac{\hbar}{\sqrt{2}}W'.
\end{equation}
There are two cases. The first is $A\psi_{0}^{(1)}\neq 0$, $E_{0}^{(1)}\neq 0$, $A^{\dagger}\psi_{0}^{(2)}\neq 0$ and $E_{0}^{(2)}\neq 0$. We have
\newline
\begin{equation}
E_{n}^{(2)}=E_{n}^{(1)}>0,\quad \psi_{n}^{(2)}=\frac{1}{\sqrt{E_{n}^{(1)}}}A\psi_{n}^{(1)},\quad  \psi_{n}^{(1)}=\frac{1}{\sqrt{E_{n}^{(2)}}}A^{\dagger}\psi_{n}^{(2)} \quad .
\end{equation}
and the two Hamiltonians are isospectral. This case corresponds to broken supersymmetry.
\newline
For the second case the supersymmetry is unbroken and we have $A\psi_{0}^{(1)}= 0$, $E_{0}^{(1)}= 0$, $A^{\dagger}\psi_{0}^{(2)}\neq 0$ and $E_{0}^{(2)}\neq 0$. Without lost of generality we take $H_{1}$ as the potential having the zero energy ground state. We have
\newline
\begin{equation}
E_{n}^{(2)}=E_{n+1}^{(1)},\quad E_{0}^{(1)}=0,\quad \psi_{n}^{(2)}=\frac{1}{\sqrt{E_{n+1}^{(1)}}}A\psi_{n+1}^{(1)},\quad  \psi_{n+1}^{(1)}=\frac{1}{\sqrt{E_{n}^{(2)}}}A^{\dagger}\psi_{n}^{(2)} \quad .
\end{equation}
We can define the matrices
\begin{equation}
H = \begin{pmatrix}H_{1} & 0 \\ 0 & H_{2}\end{pmatrix}\quad Q = \begin{pmatrix}0 & 0 \\ A & 0\end{pmatrix}\quad Q^{\dagger} = \begin{pmatrix} 0 & A^{\dagger} \\ 0 & 0\end{pmatrix}\quad .
\end{equation}
\newline
We get the relations 
\newline
\begin{equation}
[H,Q]=[H,Q^{\dagger}]=0, \quad  \{Q,Q\}=\{Q^{\dagger},Q^{\dagger}\}=0, \quad  \{Q,Q^{\dagger}\}=H \quad .
\end{equation}
\newline
We have a sl(1|1) superalgebra and $H_{1}$ and $H_{2}$ are superpartners. By construction, all our potentials can be viewed as the sum of two one dimensional potentials $H=H_{x}+H_{y}$. The unidimensional parts of the three reducible potentials  and the irreducible Potential 2 and Potential 3 are known in SUSYQM. These potentials have the shape invariance property [29,31]. We will show that Potentials 1,4,5 and 6 can be also discussed from the point of view of supersymmetry.
\newline
\subsection{Potential 1}
The Hamiltonian is
\begin{equation}
H=H_{x}+H_{y}=\frac{P_{x}^{2}}{2} + \frac{P_{y}^{2}}{2} + \hbar^{2}(
\frac{x^{2}+y^{2}}{8a^{4}} +
\frac{1}{(x-a)^{2}}+\frac{1}{(x+a)^{2}})\quad .
\end{equation}
\newline
\newline
Let us define the two operators 
\newline
\newline
\begin{equation}
b^{\dagger}= \frac{1}{\sqrt{2}}(-\hbar \frac{d}{dx} - \frac{ \hbar}{2a^{2}}x
-\hbar (\frac{-1}{x-a}+\frac{-1}{x+a}))
\end{equation}
\begin{equation}
b= \frac{1}{\sqrt{2}}(\hbar \frac{d}{dx} - \frac{
\hbar}{2a^{2}}x -\hbar (\frac{-1}{x-a}+\frac{-1}{x+a}))\quad .
\end{equation}
For $a=ia_{0},\quad a_{0}\in \mathbb{R}$ we have 
\newline
\begin{equation}
H_{1}=b^{\dagger}b=\frac{P_{x}^{2}}{2}+\frac{\hbar^{2}x^{2}}{8a_{0}^{4}}+\frac{\hbar^{2}}{(x-ia_{0})^{2}}+\frac{\hbar^{2}}{(x+ia_{0})^{2}}+\frac{3\hbar^{2}}{4a_{0}^{2}}
\end{equation}
\begin{equation}
H_{2}=bb^{\dagger}=\frac{P_{x}^{2}}{2}+\frac{\hbar^{2} x^{2}}{8 a_{0}^{4}}+
\frac{5\hbar^{2}}{4a_{0}^{2}}\quad .
\end{equation}
These two unidimensional Hamiltonians are almost isospectral. $H_{1}$ has a zero energy ground state. 
The supersymmetry is unbroken. This potential was discussed in Ref. 33. Non singular superpartners of the harmonic oscillator were discussed in Ref. 34 and 35. Coherent states of superpartners of the harmonic oscillator have also been studied [36].
Wee see that $H_{1}=H_{x}+\frac{3\hbar^{2}}{4a_{0}^{2}}$ is the Hamiltonian that we are interested in and its superpartner $H_{2}$ corresponds to a harmonic oscillator.
\newline
We apply results for the unbroken supersymmetry. The zero energy ground state satisfies $b\phi_{0}=0$ and is 
\newline
\begin{equation}
\phi_{0}(x)=a_{0}^{\frac{3}{2}}(\frac{2}{\pi})^{\frac{1}{4}}\frac{e^{\frac{-x^{2}}{4a_{0}^{2}}}}{a_{0}^{2}+x^{2}} \quad .
\end{equation}
\newline
The other eigenfunctions of $H_{1}$ are obtained by the equation $\phi_{n+1}^{(1)}=\frac{1}{\sqrt{E_{n}^{(2)}}}b^{\dagger}\phi_{n}^{(2)}$. In this case $\psi_{n}^{(2)}$ are only the eigenfunctions of the harmonic oscillator ($H_{2}$) that are written in terms of Hermite polynomials. We get directly for $H_{1}$
\newline
\newline
\begin{equation}
\phi_{k_{1}+1}(x)=b^{\dagger}(\frac{1}{\sqrt{2^{k_{1}}k_{1}!}}(\frac{1}{2a_{0}^{2}\pi})^{\frac{1}{4}}e^{\frac{-1}{4a_{0}^{2}}x^{2}}H_{k_{1}}(\sqrt{\frac{1}{2a_{0}^{2}}}x))
\end{equation}
\[=\frac{a_{0}}{\sqrt{(k_{1}+3)}}(\frac{1}{2a_{0}^{2}\pi})^{\frac{1}{4}}\frac{1}{\sqrt{2^{k_{1}}k_{1}!}}e^{-\frac{x^{2}}{4a_{0}^{2}}}(\frac{(x^{3}+3xa_{0}^{2})}{a_{0}^{2}(x^{2}+a_{0}^{2})}H_{k_{1}}-\frac{2k_{1}}{\sqrt{2}a_{0}}\lambda H_{k_{1}-1}),         \]
$\lambda$=1 for $k_{1}\geq 1$, $\lambda$=0 for $k_{1}$=0. With this expression we get for $k_{1}=0$
\begin{equation}
\phi_{1}=\frac{1}{\sqrt{3}(2\pi)^{\frac{1}{4}}a_{0}^{\frac{3}{2}}}e^{-\frac{x^{2}}{4a_{0}^{2}}}\frac{x(3a_{0}^{2}+x^{2})}{a_{0}^{2}+x^{2}}\quad .
\end{equation}
\newline
\newline
We have the following energy spectrum for $H_{1}$
\begin{equation}
E_{0}^{(1)}=0, \quad E_{k_{1}+1}^{(1)}=\frac{\hbar^{2}}{2a_{0}^{2}}(k_{1}+3)\quad .
\end{equation}
\newline
\newline
We thus obtain the spectrum of $H_{x}$ ( the x part of the irreducible Potential 1):
\begin{equation}
E_{0}^{x}=-\frac{3\hbar^{2}}{4a_{0}^{2}},\quad E_{k_{1}+1}^{x}=\frac{\hbar^{2}}{2a_{0}^{2}}(k_{1}+\frac{3}{2})\quad .
\end{equation}
\newline
\newline
If we add $H_{y}$ to these results we get the energy spectrum and the eigenfunctions of the Potential 1. There are two families of solutions. The first corresponds to the energies
\newline
\begin{equation}
E=\frac{(k_{1}+k_{2}+2)\hbar^{2}}{2a_{0}^{2}}=\frac{(p+2)\hbar^{2}}{2a_{0}^{2}}
\end{equation}
with eigenfunctions
\newline
\begin{equation}
\phi_{k_{1}+1}(x)=\frac{a_{0}}{\sqrt{(k_{1}+3)}}(\frac{1}{2a_{0}^{2}\pi})^{\frac{1}{4}}\frac{1}{\sqrt{2^{k_{1}}k_{1}!}}e^{-\frac{x^{2}}{4a_{0}^{2}}}(\frac{(x^{3}+3xa_{0}^{2})}{a_{0}^{2}(x^{2}+a_{0}^{2})}H_{k_{1}}-\frac{2k_{1}}{\sqrt{2}a_{0}}\lambda H_{k_{1}-1})
\end{equation}
\begin{equation}
\chi_{k_{2}}(y)=\frac{1}{\sqrt{2^{k_{2}}k_{2}!}}(\frac{1}{2a_{0}^{2}\pi})^{\frac{1}{4}}e^{\frac{-1}{4a_{0}^{2}}y^{2}}H_{k_{2}}(\sqrt{\frac{1}{2a_{0}^{2}}}y)
\end{equation}
\newline
and is also obtained from the cubic algebra. The second corresponds to the energies
\begin{equation}
E=\frac{\hbar^{2}(k_{2}-1) }{2a_{0}^{2}}
\end{equation}
with the corresponding eigenfunctions
\newline
\begin{equation}
\psi(x,y)=\phi_{0}(x)\chi_{k_{2}}(y),\quad \phi_{0}(x)=a_{0}^{\frac{3}{2}}(\frac{2}{\pi})^{\frac{1}{4}}\frac{e^{\frac{-x^{2}}{4a_{0}^{2}}}}{a_{0}^{2}+x^{2}}
\end{equation}
and $\chi_{k_{2}}(y)$ as in Eq.(4.19).
\newline
The two states obtained from Eq.(3.8) are given by Eq.(4.20) for $k_{2}$=0,1. For $k_{3} \geq 3$ there are common eigenvalues given by Eq.(4.17) and Eq.(4.20) and therefore the degeneracy is p+2.
\newline
Let us consider the case $a \in \mathbb{R}$. We have the following Hamiltonians
\newline
\begin{equation}
H_{1}=b^{t}b=\frac{P_{x}^{2}}{2}+\frac{\hbar^{2}x^{2}}{8a^{4}}+\frac{\hbar^{2}}{(x-a)^{2}}+\frac{\hbar^{2}}{(x+a)^{2}}-\frac{3\hbar^{2}}{4a^{2}}
\end{equation}
\begin{equation}
H_{2}=bb^{t}=\frac{P_{x}^{2}}{2}+\frac{\hbar^{2} x^{2}}{8 a^{4}}-
\frac{5\hbar^{2}}{4a^{2}} \quad .
\end{equation}
\newline
This case is more complicated because of the singularities on the x-axis for the Hamiltonian $H_{1}$. We have a regular Hamiltonian connected to a singular one and we have also for $H_{2}$ negative energy states. Such situations have attracted a lot of attention and many articles were devoted to such singular potentials. An important case is the one of Jevicki and Rodrigues [37,38]. The corresponding Hamiltonians are
\newline
\begin{equation}
\quad H_{-}=\frac{d^{2}}{dx^{2}}+x^{2}-3,\quad H_{+}=-\frac{d^{2}}{dx^{2}}+x^{2}+\frac{2}{x^{2}}-1.
\end{equation}
\newline
Factorization of Hamiltonians $H_{1}$ and $H_{2}$ given by Eq.(4.22) and (4.23) give us an algebraic relation that does not take into account the presence of singularities or boundary conditions. The wavefunctions given in Eq.(4.3) and (4.4) do not necessarily belong to the Hilbert space of square integrable functions. The potential in Eq.(4.22) has impenetrable barriers coming from the singularities. We can consider the superpartner to be the harmonic oscillator with two infinite barriers (at $x=\pm a$) to recover the supersymmetry [39]. In the Ref. 39 a superpartner of the harmonic oscillator with one singularity was considered but the method can be extended to more singularities. The only case that was solved analytically and where the energy levels are equidistant is when the singularity was at the origin. In our case we were not able to solve analytically and we leave for future investigations these numerical calculations that appear interesting from a phenomenological point of view. Singular potentials were also investigated by A.Das and S.A.Pernice [40] by means of the regularization method. M.Znojil [41] has discussed another method that consist in the complexification of the potential. In Section 6 we will discuss the complexification of the irreducible quantum superintegrable Potential 1.
\newline
\subsection{Potential 4}
We apply these results to the next irreducible potential
\newline
\begin{equation}
V = \hbar^{2}(\frac{9x^{2}+y^{2}}{8a^{4}} + \frac{1}{(y-a)^{2}}+\frac{1}{(y+a)^{2}})
\end{equation}
We can also use SUSYQM because the y part is the same as the x part of Potential 1. For the case $ a=ia_{0},\quad a_{0}\in \mathbb{R}$ we find with energy 
\begin{equation}
E=\frac{\hbar^{2}}{2a_{0}^{2}}(3k_{1}+k_{2}+3)
\end{equation}
with the corresponding eigenfunctions
\newline
\begin{equation}
\chi_{k_{1}}(x)=\frac{1}{\sqrt{2^{k_{1}}k_{1}!}}(\frac{3}{2a_{0}^{2}\pi})^{\frac{1}{4}}e^{-\frac{3}{4a_{0}^{2}}x^{2}}H_{k_{1}}(\sqrt{\frac{3}{2a_{0}^{2}}}x)
\end{equation}
\begin{equation}
\phi_{k_{2}+1}(y)=\frac{a_{0}}{\sqrt{(k_{2}+3)}}(\frac{1}{2a_{0}^{2}\pi})^{\frac{1}{4}}\frac{1}{\sqrt{2^{k_{2}}k_{2}!}}e^{-\frac{y^{2}}{4a_{0}^{2}}}(\frac{(y^{3}+3ya_{0}^{2})}{a_{0}^{2}(y^{2}+a_{0}^{2})}H_{k_{2}}-\frac{2k_{2}}{\sqrt{2}a_{0}}\lambda H_{k_{2}-1})
\end{equation}
\newline
\newline
and we get from the singlet state the energies
\begin{equation}
E=\frac{\hbar^{2}}{2a_{0}^{2}}(3k_{1})
\end{equation}
and eigenfunctions
\begin{equation}
\psi(x,y)=\phi_{0}(y)\chi_{k_{1}}(x),\quad \phi_{0}(y)=a_{0}^{\frac{3}{2}}(\frac{2}{\pi})^{\frac{1}{4}}\frac{e^\frac{-y^{2}}{4a_{0}^{2}}}{a_{0}^{2}+y^{2}}
\end{equation}
and $\chi_{k_{1}}(x)$ as in Eq.(4.27).
\newline
\subsection{Potential 5}
The potential is
\begin{equation}
V = \hbar^{2}[\frac{1}{8a^{4}}[(x^{2}+y^{2})+\frac{1}{y^{2}}+\frac{1}{(x+a)^{2}}+\frac{1}{(x-a)^{2}}]
\end{equation}
For the case $a=ia_{0},\quad a_{0}\in \mathbb{R}$ we have
\begin{equation}
E=\frac{(k_{1}+ 2k_{2}+5)}{2a_{0}^{2}}\hbar^{2}
\end{equation}
with the eigenfunctions given by
\newline
\begin{equation}
\phi_{k_{1}}(x)=\frac{a_{0}}{\sqrt{(k_{1}+3)}}(\frac{1}{2a_{0}^{2}\pi})^{\frac{1}{4}}\frac{1}{\sqrt{2^{k_{1}}k_{1}!}}e^{-\frac{x^{2}}{4a_{0}^{2}}}(\frac{(x^{3}+3xa_{0}^{2})}{a_{0}^{2}(x^{2}+a_{0}^{2})}H_{k_{1}}-\frac{2k_{1}}{\sqrt{2}a_{0}}\lambda H_{k_{1}-1})
\end{equation}
\begin{equation}
\chi_{k_{2}}(y)=(\frac{1}{2a_{0}^{2}})^{\frac{1}{4}}(\frac{k_{2}!(\frac{1}{2a_{0}^{2}}
)^{\frac{3}{2}}}{\Gamma(k_{2}+\frac{5}{2})})^{\frac{1}{2}}e^{\frac{-y^{2}}{4a_{0}^{2}}
}y^{2}L_{k_{2}}^{\frac{3}{2}}(\frac{y^{2}}{2a_{0}^{2}}
)
\end{equation}
where $L_{k}^{\nu}(z)$ are Laguerre polynomials
\newline
We have also the energies
\begin{equation}
E=\frac{\hbar^{2}(2k_{2}+ 2)}{2a_{0}^{2}}
\end{equation}
with the corresponding eigenfunctions
\newline
\begin{equation}
\psi(x,y)=\chi_{k_{2}}(y)\phi_{0}(x),\quad \phi_{0}(x)=a_{0}^{\frac{3}{2}}(\frac{2}{\pi})^{\frac{1}{4}}\frac{e^{\frac{-x^{2}}{4a_{0}^{2}}}}{a_{0}^{2}+x^{2}}
\end{equation}
and $\chi_{k_{2}}(y)$ as Eq.(4.34).
\subsection{Potential 6}
We consider the potential
\begin{equation}
V = \hbar^{2}[\frac{1}{8a^{4}}(x^{2}+y^{2})+\frac{1}{(y+a)^{2}}+\frac{1}{(y-a)^{2}}
+\frac{1}{(x+a)^{2}}+\frac{1}{(x-a)^{2}}]
\end{equation}
\newline
\newline
For the case $a=ia_{0},\quad a_{0}\in \mathbb{R}$ we have the energies
\begin{equation}
E=\frac{(k_{1}+k_{2}+3)}{2a_{0}^{2}}\hbar^{2}
\end{equation}
with the eigenfunctions given by 
\newline
\begin{equation}
\phi_{k_{1}+1}(x)=\frac{a_{0}}{\sqrt{(k_{1}+3)}}(\frac{1}{2a_{0}^{2}\pi})^{\frac{1}{4}}\frac{1}{\sqrt{2^{k_{1}}k_{1}!}}e^{-\frac{x^{2}}{4a_{0}^{2}}}(\frac{(x^{3}+3xa_{0}^{2})}{a_{0}^{2}(x^{2}+a_{0}^{2})}H_{k_{1}}-\frac{2k_{1}}{\sqrt{2}a_{0}}\lambda H_{k_{1}-1})
\end{equation}
\begin{equation}
\chi_{k_{2}+1}(y)=\frac{a_{0}}{\sqrt{(k_{2}+3)}}(\frac{1}{2a_{0}^{2}\pi})^{\frac{1}{4}}\frac{1}{\sqrt{2^{k_{2}}k_{2}!}}e^{-\frac{y^{2}}{4a_{0}^{2}}}(\frac{(y^{3}+3ya_{0}^{2})}{a_{0}^{2}(y^{2}+a_{0}^{2})}H_{k_{2}}-\frac{2k_{2}}{\sqrt{2}a_{0}}\lambda H_{k_{2}-1})
\end{equation}
\newline
The singlet state in the x part of the Hamiltonian gives the energies
\begin{equation}
E=\frac{\hbar^{2}(k_{2})}{2a_{0}^{2}}
\end{equation}
with eigenfunctions  
\begin{equation}
\psi(x,y)=\chi_{k_{2}}(y)\phi_{0}(x),\quad \phi_{0}(x)=a_{0}^{\frac{3}{2}}(\frac{2}{\pi})^{\frac{1}{4}}\frac{e^{\frac{-x^{2}}{4a_{0}^{2}}}}{a_{0}^{2}+x^{2}}
\end{equation}
and $\chi_{k_{2}}(y)$ as Eq.(4.40).
\newline
We also obtain another kind of solution from the singlet state in the y part. The energies are 
\begin{equation}
E=\frac{\hbar^{2}(k_{1})}{2a_{0}^{2}}
\end{equation}
with the corresponding eigenfunctions
\begin{equation}
\psi(x,y)=\phi_{k_{1}}(x)\chi_{0}(y),\quad \chi_{0}(y)=a_{0}^{\frac{3}{2}}(\frac{2}{\pi})^{\frac{1}{4}}\frac{e^{\frac{-y^{2}}{4a_{0}^{2}}}}{a_{0}^{2}+y^{2}}
\end{equation}
and $\phi_{k_{1}}(x)$ as Eq.(4.39).
\newline
and a state coming from the singlet state in the two parts with energies 
\begin{equation}
E=-\frac{3\hbar^{2}}{2a_{0}^{2}}
\end{equation}
with the following expresion for the eigenfunctions
\begin{equation}
\phi_{0}(x)=a_{0}^{\frac{3}{2}}(\frac{2}{\pi})^{\frac{1}{4}}\frac{e^{\frac{-x^{2}}{4a_{0}^{2}}}}{a_{0}^{2}+x^{2}}
,\quad \chi_{0}(y)=a_{0}^{\frac{3}{2}}(\frac{2}{\pi})^{\frac{1}{4}}\frac{e^{\frac{-y^{2}}{4a_{0}^{2}}}}{a_{0}^{2}+y^{2}}.
\end{equation}
\section{Generating spectrum algebra}
The supersymmetry allows us to find the creation and annihilation operators of the x part of the irreducible Potential 1. They are given by
\begin{equation}
M=b^{\dagger}cb,\quad M^{\dagger}=b^{\dagger}c^{\dagger}b
\end{equation}
where c and $c^{\dagger}$ are annihilation and creation operators of the superpartner $H_{2}$ that is a harmonic oscillator. We have
\begin{equation}
c=\frac{\hbar}{2a^{2}}(x+2a^{2}\frac{d}{dx}),\quad c^{\dagger}=\frac{\hbar}{2a^{2}}(x-2a^{2}\frac{d}{dx})
\end{equation}
and
\begin{equation}
M=\frac{1}{\sqrt{2}}(-\hbar \frac{d}{dx} - \frac{ \hbar}{2a^{2}}x
+\hbar (\frac{1}{x-a}+\frac{1}{x+a}))\frac{\hbar}{2a^{2}}(x+2a^{2}\frac{d}{dx})\frac{1}{\sqrt{2}}(\hbar \frac{d}{dx} - \frac{ \hbar}{2a^{2}}x +\hbar (\frac{1}{x-a}+\frac{1}{x+a}))
\end{equation}
\begin{equation}
M^{\dagger}=\frac{1}{\sqrt{2}}(-\hbar \frac{d}{dx} - \frac{ \hbar}{2a^{2}}x
+\hbar (\frac{1}{x-a}+\frac{1}{x+a}))\frac{\hbar}{2a^{2}}(x-2a^{2}\frac{d}{dx})\frac{1}{\sqrt{2}}(\hbar \frac{d}{dx} - \frac{ \hbar}{2a^{2}}x +\hbar (\frac{1}{x-a}+\frac{1}{x+a}))\quad .
\end{equation}
The zero energy ground state given by the Eq.(4.12) is annihilated be the annihilation operator but also by the creation operator.
\newline
\begin{equation}
M\phi_{0}(x)=M^{\dagger}\phi_{0}=0
\end{equation}
The creation and annihilation operator for the y part ($H_{y}$) of the Potential 1 are
\begin{equation}
L=\frac{\hbar}{2a^{2}}(y+2a^{2}\frac{d}{dy}), \quad L^{\dagger}=\frac{\hbar}{2a^{2}}(y-2a^{2}\frac{d}{dy})\quad .
\end{equation}
We have the commutators
\begin{equation}
[M,M^{\dagger}]=\frac{3}{4}(H+\frac{1}{2}A)^{2}-\frac{\hbar^{2}}{a^{2}}(H+\frac{1}{2}A)-\frac{3\hbar^{4}}{16a^{4}},\quad [L,L^{\dagger}]=1 \quad .
\end{equation}
We consider the following operators [16]
\begin{equation}
E_{+}=M^{\dagger}L^{\dagger},\quad E_{-}=ML,\quad F_{+}=(M^{\dagger})^{2},\quad F_{-}=M^{2},\quad G_{+}=(L^{\dagger})^{2},\quad G_{-}=L^{2} .
\end{equation}
We add to these operators the Hamiltonian, the integrals of motion A, B and C (Eq.(3.1) (3.2) and (3.3)). 
We have the following quintic algebra that contains 45 relations where the cubic algebra appears as a subalgebra:
\newline
\begin{equation}
\end{equation}
\[ [H,A]=0,\quad [H,B]=0,\quad [H,C]=0,\quad [H,E_{\pm}]=\pm (\frac{\hbar^{2}}{a^{2}})E_{\pm}, \]
\[ [H,F_{\pm}]=\pm (\frac{\hbar^{2}}{a^{2}})F_{\pm},\quad [H,G_{\pm}]=\pm (\frac{\hbar^{2}}{a^{2}})G_{\pm},\]
\[ [A,B]=C,\quad [A,C]=\frac{4h^{4}}{a^{4}}B, \quad [A,E_{\pm}]=0,\quad [A,F_{\pm}]=\pm (\frac{2\hbar^{2}}{a^{2}})F_{\pm},\]
\[ [A,G_{\pm}]=\mp (\frac{2\hbar^{2}}{a^{2}})G_{\pm},\quad [B,C]= -2\hbar^{2}A^{3} - 6\hbar^{2}A^{2}H \]
\[+ 8\hbar^{2}H^{3} + 6\frac{\hbar^{4}}{a^{2}}A^{2} + 8\frac{\hbar^{4}}{a^{2}}HA - 8\frac{\hbar^{4}}{a^{2}}H^{2} + 2\frac{\hbar^{6}}{a^{4}}A - 2\frac{\hbar^{6}}{a^{4}}H - 6\frac{\hbar^{8}}{a^{6}}, \]
\[ [B,E_{-}]= -2i\hbar F_{-}+\frac{3i\hbar}{2}(H+\frac{1}{2}A)^{2}G_{-} -\frac{2i\hbar^{3}}{a^{2}}(H+\frac{1}{2}A)G_{-}-\frac{3i\hbar^{5}}{8a^{4}}G_{-},\]
\[ [B,E_{+}]= -2i\hbar F_{+}+\frac{3i\hbar}{2}(H+\frac{1}{2}A)^{2}G_{+} -\frac{2i\hbar^{3}}{a^{2}}(H+\frac{1}{2}A)G_{+}-\frac{3i\hbar^{5}}{8a^{4}}G_{+},\]
\[ [B,F_{-}]= 3i\hbar(H+\frac{1}{2}A)^{2}E_{-} -\frac{7i\hbar^{3}}{a^{2}}(H+\frac{1}{2}A)E_{-}+\frac{11i\hbar^{5}}{4a^{4}}E_{-},\]
\[ [B,F_{+}]= 3i\hbar(H+\frac{1}{2}A)^{2}E_{+} -\frac{i\hbar^{3}}{a^{2}}(H+\frac{1}{2}A)E_{+}-\frac{5i\hbar^{5}}{4a^{4}}E_{+},\]
\[ [B,G_{+}]=-4i\hbar E_{+},\quad [B,G_{-}]=-4i\hbar E_{-},  \]
\[ [C,E_{-}]=\frac{4i\hbar^{3}}{a^{3}} F_{-}+\frac{3i\hbar^{3}}{a^{2}}(H+\frac{1}{2}A)^{2}G_{-} -\frac{4i\hbar^{5}}{a^{4}}(H+\frac{1}{2}A)G_{-}-\frac{3i\hbar^{7}}{4a^{6}}G_{-},\]
\[ [C,E_{+}]=\frac{-4i\hbar^{3}}{a^{3}} F_{+}-\frac{3i\hbar^{3}}{a^{2}}(H+\frac{1}{2}A)^{2}G_{+} +\frac{4i\hbar^{5}}{a^{4}}(H+\frac{1}{2}A)G_{+}+\frac{3i\hbar^{7}}{4a^{6}}G_{+},\]
\[ [C,F_{-}]= \frac{6i\hbar^{3}}{a^{2}}(H+\frac{1}{2}A)^{2}E_{-} -\frac{14i\hbar^{5}}{a^{4}}(H+\frac{1}{2}A)E_{-}+\frac{11i\hbar^{7}}{2a^{6}}E_{-},\]
\[ [C,F_{+}]= -\frac{6i\hbar^{3}}{a^{2}}(H+\frac{1}{2}A)^{2}E_{+} +\frac{2i\hbar^{5}}{a^{4}}(H+\frac{1}{2}A)E_{+}+\frac{5i\hbar^{7}}{2a^{6}}E_{+},\]
\[ [C,G_{\pm}]=\mp\frac{8i\hbar^{3}}{a^{2}}E_{\pm},\quad [E_{\pm},F_{\pm}]=0,\quad [E_{\pm},G_{\pm}]=0, \]
\[ [E_{-},E_{+}]=\frac{-a^{2}\hbar^{2}}{16}A^{3}+\frac{3a^{2}\hbar^{2}}{4}AH^{2}+a^{2}\hbar^{2}H^{3}+\frac{\hbar^{4}}{8}A^{2} \]
\[-\frac{\hbar^{4}}{2}AH-\frac{3\hbar^{4}}{2}H^{2}+\frac{\hbar^{6}}{16a^{2}}A-\frac{\hbar^{6}}{4a^{2}}H+\frac{3\hbar^{8}}{8a^{4}}, \]
\[ [E_{+},F_{-}] =\frac{-3ia^{2}\hbar}{16}C(H+\frac{1}{2}A)^{2}+\frac{3i\hbar^{3}}{8}B(H+\frac{1}{2}A)^{2}+\frac{7i\hbar^{3}}{16}C(H+\frac{1}{2}A)\]
\[-\frac{7i\hbar^{5}}{8a^{2}}B(H+\frac{1}{2})-\frac{11i\hbar^{5}}{64a^{2}}C+\frac{11i\hbar^{7}}{32a^{4}}B,\]
\[ [E_{-},F_{+}] =\frac{3ia^{2}\hbar}{16}C(H+\frac{1}{2}A)^{2}+\frac{3i\hbar^{3}}{8}B(H+\frac{1}{2}A)^{2}-\frac{i\hbar^{3}}{16}C(H+\frac{1}{2}A)\]
\[-\frac{i\hbar^{5}}{8a^{2}}B(H+\frac{1}{2})-\frac{5i\hbar^{5}}{64a^{2}}C-\frac{5i\hbar^{7}}{32a^{4}}B,\]
\[ [E_{-},G_{+}]=\frac{ia^{2}\hbar}{4}C-\frac{i\hbar^{3}}{2}B,\quad [E_{+},G_{-}]=\frac{-ia^{2}\hbar}{4}C-\frac{i\hbar^{3}}{2}B,\]
\[ [F_{-},F_{+}]= \frac{3a^{2}\hbar^{2}}{4}(H+\frac{1}{2}A)^{5}-\frac{5\hbar^{4}}{2}(H+\frac{1}{2}A)^{4}+\frac{25\hbar^{6}}{8a^{2}}(H+\frac{1}{2}A)^{3}\]
\[-\frac{5\hbar^{8}}{4a^{4}}(H+\frac{1}{2}A)^{2}-\frac{53\hbar^{10}}{64a^{6}}(H+\frac{1}{2}A)+\frac{15\hbar^{12}}{32a^{8}},\]
\[ [F_{\pm},G_{\pm}]=[F_{\pm},G_{\mp}]=0,\quad [G_{-},G_{+}]=4a^{2}\hbar^{2}(H+\frac{1}{2}A).\]
This polynomial algebra is the spectrum-generating algebra.
\section{Complexification of superintegrable potentials}
In quantum mechanics textbooks the Hermiticity of the Hamiltonian is often presented as a condition for the energy spectrum to be real. There exist other requirements that can be chosen without loosing essential features of quantum mechanics. One requirement that appears more physical is the space-time reflection symmetry i.e. the Hamiltonian is invariant under the PT transformation [42], i.e. the simultaneous reflections P: x$\rightarrow$-x,  p$\rightarrow$-p and $\tau$: x$\rightarrow$x   p$\rightarrow$-p, i$\rightarrow$-i. For potentials invariant under such transformations the energy spectrum can also consist of complex-conjugate pairs of eigenvalues. The PT-symmetry is thus said to be broken. The notion of Pseudo-Hermiticity was introduced by A.Mostafazadeh [43]. He shows also that every Hamiltonian with a real spectrum is pseudo-Hermitian and that all PT-symmetric Hamiltonians studied belong to the class of pseudo-Hermitian Hamiltonian. The replacement of the condition that the Hamiltonian is Hermitian by a weaker condition allows us to study many new kinds of Hamiltonians that would have been excluded and from a phenomenological point of view may describe physic phenomena. The case $H=p^{2}+x^{2}(ix)^{\delta}$ was studied in detail by C.Bender in 1998 [42]. 
\newline
Complexification has been proposed as a natural way to regularize singular potentials [41]. It consists in a transformation of the type x $\rightarrow$ x - i$\epsilon$ applied to a potential. The harmonic oscillator and the Smorodinsky-Winternitz potential are PT-symmetric Hamiltonian after a complexification [41].
\newline
We will consider the complexification of the Hamiltonian
\begin{equation}
H =H_{x}+H_{y}= \frac{P_{x}^{2}}{2} + \frac{P_{y}^{2}}{2} + \hbar^{2}(
\frac{(x-i\epsilon)^{2}+(y-i\epsilon)^{2}}{8a^{4}} +
\frac{1}{(x-i\epsilon-a)^{2}}+\frac{1}{(x-i\epsilon+a)^{2}}).
\end{equation}
\newline
The complex harmonic oscillator Hamiltonian $H_{y}$ is known to be PT-symmetric. It's energy spectrum is real, namely:
\begin{equation}
E=\frac{\hbar^{2}}{2a^{2}}(m+\frac{1}{2})\quad .
\end{equation}
The eigenfunctions are
\newline
\newline
\begin{equation}
\phi_{m}(y)=N_{m}e^{-\frac{(y-i\epsilon)^{2}}{4a^{2}}}H_{m}(\frac{(y-i\epsilon)}{\sqrt{2}a})
\end{equation}
(here and below $N_{m}$ is a normalization constant).
\newline
To get the energy spectrum and the eigenfunctions of $H_{x}$ we complexify the operators given by the Eq.(4.8) and (4.9). We get two PT-symmetric Hamiltonians. This transformation allows to regularize $H_{x}$ when $a \in \mathbb{R}$. The (real) energy levels and eigenfunctions of the Hamiltonian $H_{2}$ are known.
\begin{equation}
H_{1}=b^{'}b=\frac{P_{x}^{2}}{2}+\frac{\hbar^{2}(x-i\epsilon)^{2}}{8a^{4}}+\frac{\hbar^{2}}{(x-i\epsilon-a)^{2}}+\frac{\hbar^{2}}{(x-i\epsilon+a)^{2}}-\frac{3\hbar^{2}}{4a^{2}}
\end{equation}
\begin{equation}
H_{2}=bb^{'}=\frac{P_{x}^{2}}{2}+\frac{\hbar^{2} (x-i\epsilon)^{2}}{8 a^{4}}-
\frac{5\hbar^{2}}{4a^{2}}
\end{equation}
\newline 
The Darboux transformation is still valid for non-Hermitian Hamiltonians but supersymmetry is replaced by pseudo-supersymmetry [44].
We have  $b^{'}\psi_{gr}=0$ that correspond to the zero energy state of $H_{2}$
\newline
\begin{equation}
\psi_{gr}=N_{gr}e^{-\frac{(x-i\epsilon)^{2}}{4a^{2}}}(a^{2}-(x-i\epsilon)^{2})
\end{equation}
We can obtain the eigenfunction of $H_{1}$ by applying $b^{'}$ on other state of $H_{2}$ given in terms of Hermite polynomials. We get
\newline
\begin{equation}
\phi_{n}=N_{n}e^{\frac{-(x-i\epsilon)^{2}}{4a^{2}}}(\frac{2(x-i\epsilon)}{(x-i\epsilon)^{2}-a^{2}}H_{n+3}(\frac{(x-i\epsilon)}{\sqrt{2}a})
\end{equation}
\[-\frac{2(n+3)}{\sqrt{2}a}H_{n+2}(\frac{(x-i\epsilon)}{\sqrt{2}a}))\]
\newline
Let us give the explicit expression for the ground state and the first excited state
\newline
\begin{equation}
\phi_{0}=N_{0}e^{\frac{-(x-i\epsilon)^{2}}{4a^{2}}}\frac{(3a^{4}+(x-i\epsilon)^{4})}{(a^{2}-(x-i\epsilon)^{2})}
\end{equation}
\begin{equation}
\phi_{1}=N_{1}e^{\frac{-(x-i\epsilon)^{2}}{4a^{2}}}\frac{(3a^{4}+2a^{2}i(x-i\epsilon)+(x-i\epsilon)^{4})(x-i\epsilon)   }{(a^{2}-(x-i\epsilon)^{2})}
\end{equation}
\newline
The probabilistic interpretation of the wave function of non-Hermitian quantum systems [45] is given by a pseudo-norm that is not positive definite.
\newline
\begin{equation}
\int_{-\infty}^{\infty}dx\psi^{*}(-x)\psi(x)=\sigma,\quad \sigma=\pm 1
\end{equation}
\newline
The corresponding energy spectrum is given by
\newline
\begin{equation}
E_{n}=\frac{(n+1)\hbar^{2}}{2a^{2}}
\end{equation}
\newline
We obtain for the complexified superintegrable potential the energy spectrum
\newline
\begin{equation}
E=\frac{(n+m+3)\hbar^{2}}{2a^{2}}=\frac{(p+3)\hbar^{2}}{2a^{2}}
\end{equation}
\newline
with eigenfunction given by Eq.(6.3) and (6.7).
\section{Conclusion}
The main result of this article is that we have constructed a Fock type representation for the most general cubic algebra generated by a second order and a third order order integral of motion by the means of parafermionic algebras. We present in detail the cubic algebra for all irreducible quantum superintegrable potentials, the unitary representations and the corresponding energy spectra. All cases with finite cubic algebras belong to Case 2 of Section 2. Thus they correspond to $\beta=0$ in Eq.(2.5) and the structure function is given by Eq. (2.22).  In two cases of irreducible potentials the integrals of motion do not close in a finite dimensional cubic algebra. It could be interesting to see what kind algebraic structure is involved in these cases. Comparing with a earlier article [10] we can see from this article how the cubic Poisson algebra is deformed into a cubic algebra in quantum mechanics. 
\newline
The method that we use to find energy spectra with the cubic algebra is independant of the choice of coordinate systems. We could apply these results in the future to systems with a third order integral that are separable in polar, elliptic or parabolic coordinates. The method is also independant of the metric and could be applied to superintegrable systems in other spaces. The methods developed in this article could be applied to other physical systems. One such system is a Schr\"odinger equation with a position dependent mass [32], others arise in the context of supersymmetric quantum mechanics.
\newline
The Potential 3 is also a special case of the following potential [46,47]
\newline
\begin{equation}
V=\frac{\omega^{2}}{2}(k^{2}x^{2}+m^{2}y^{2})+\frac{\lambda_{1}}{x^{2}}+\frac{\lambda_{2}}{y^{2}}
\end{equation}
\newline
In general this system has integrals of motion of order greater than 3 and the more complicated polynomial algebra should be studied.
\newline
All the potentials considered in this article can also be viewed as the sum of two one-dimensional potentials $H=H_{x}+H_{y}$. We have investigated each of these unidimensional potentials in terms of supersymmetric quantum mechanics. The superintegrability of these two-dimensional potentials seems to be related to the supersymmetry property. Using the supersymmetry we have obtained the energy spectra and the eigenfunctions. We have compared the results with those obtained using the cubic algebras. One particular feature is the appearance of singlet states. For the Potential 1 there is an additional degeneracy that is not obtained by the algebraic method using the cubic algebra.
\newline
It was shown that many well known potentials such Dirac delta and Poschl-Teller display a hidden SUSY where the reflection (parity) operator play the role of the grading operator [48]. Potentials with elliptic functions can also be discussed from this point of view [49]. Potentials with elliptic functions appear in Ref. 9. These cases are not truly superintegrable since there exists a syzygy between the Hamiltonian, second order integral and the third order integral of motion but it has been shown that the third order integral can be used to obtain the eigenfunctions and the spectrum [50]. We leave quantum potentials involving Painlev\'e transcendents for a future article.
\newline
Superintegrable potentials and their integrals of motion can be complexified and investigated from the point of view of PT-symmetric quantum mechanics. The complexification appears also as a natural way to regularize the singular potentials. 
\newline
It's would be interesting to investigate the relation of pseudo-Hermitian Hamiltonians and supersymmetry with superintegrable systems. 
\newline
\newline
\textbf{Acknowledgments} The research of I.M. is supported by a doctoral
research scholarship from FQRNT of Quebec. The author thanks P.Winternitz for very helpful comments and discussions. This article was written in part while he was attending the conference Superintegrable Systems in Classical and Quantum Mechanics-Prague 2008 at the Czech Technical University of Prague. He thanks the Doppler Institute and the Department of Physics of the Faculty of Nuclear Sciences and Physical Enginnering for hospitality and the research plan MSM6840770039 of the Ministry of Education of the Czech Republic for financial support during the conference.
\section{Appendix}
\textbf{Structure function for the case $\beta \neq 0$}
\begin{equation*}
(A1)\quad \quad \Phi(N)=384\mu\beta^{10}N^{10}-1920\mu \beta^{10}N^{9}+(-1536\delta \mu
\beta^{8}+1024\nu \beta^{9}+3040\mu
\beta^{10}
\end{equation*}
\[-2304\beta^{8}\alpha^{2})N^{8}+(6144\delta \mu \beta^{8}-4096\nu \beta^{9}-640\mu
\beta^{10}+6144\beta^{8}\alpha^{2})N^{7}+ (2304\delta^{2}\mu
\beta^{6}-3072\nu \delta \beta^{7}   \]
\[ 3072\xi \beta^{8}-7680\delta \mu \beta^{8}+5120\nu \beta^{9}-2512\mu
\beta^{10}-3072\delta\beta^{6}\alpha^{2}+2304\beta^{8}\alpha^{2}+3072\beta^{7}\alpha
\gamma)N^{6}  \]
\[+(-6912\delta^{2}\mu\beta^{6}+9216\nu\delta
\beta^{7}-9216\xi\beta^{8}+1536\delta
\mu\beta^{8}-1024\nu\beta^{9}+1712\mu\beta^{10}+9216\delta\beta^{6}\alpha^{2}
   \]
\[-7680\beta^{8}\alpha^{2}-9216\beta^{7}\alpha \gamma
)N^{5}+(-1536\delta^{3}\mu\beta^{4}+3072\nu\delta^{2}\beta^{5}-6144\xi\delta
\beta^{6}+6336\delta^{2}\mu\beta^{6}  \]
\[-8448\nu\delta\beta^{7}\alpha
\gamma+8448\xi\beta^{8}+3264\delta\mu\beta^{8}-2176\nu\beta^{9}+428\mu\beta^{10}\]
\[+4608\delta^{2}\beta^{4}\alpha^{2}-8448\delta \beta^{6}\alpha^{2}+672\beta^{8}\alpha^{2}-9216\delta\beta^{5}\alpha\gamma +8448\beta^{7}     \]
\[+3072\beta^{6}\gamma^{2}+6144\beta^{6}\alpha
\epsilon+12288\beta^{7}\zeta)N^{4}+(3072\delta^{3}\mu\beta^{4}-6144\nu\delta^{2}\beta^{5}
  \]
\[+12288\xi\delta\beta^{6}-1152\delta^{2}\mu\beta^{6}+1536\nu\delta
\beta^{7}-1536\xi\beta^{8}-1920\delta
\mu\beta^{8}+1280\nu\beta^{9}-616\mu\beta^{10}-12288\delta^{2}\beta^{4}\alpha^{2}\]
\[+1536\delta\beta^{6}\alpha^{2}+2688\beta^{8}\alpha^{2}+24576\delta
\beta^{5}\alpha \gamma
-1536\beta^{7}\alpha\gamma-6144\beta^{6}\gamma^{2}-24576\beta^{6}\alpha
\epsilon-24576\beta^{7}\zeta)N^{3}\]
\[+(384\delta^{4}\mu\beta^{2}-1024\nu\delta^{3}\beta^{3}+3072\xi\delta^{2}\beta^{4}-1792\delta^{3}\mu\beta^{4}+3584\nu\delta^{2}\beta^{5}-9216\xi\delta\beta^{6}-784\delta^{2}\mu\beta^{6}-1728\xi\beta^{8}\]
\[+1728\nu\delta\beta^{7}-96\delta
\mu\beta^{8}+64\nu\beta^{9}+\frac{119}{2}\mu\beta^{10}-12288\beta^{6}K-3072\delta^{3}\beta^{2}\alpha^{2}+6912\delta^{2}\beta^{4}\alpha^{2}+1728\delta\]
\[\beta^{6}\alpha^{2}-624\beta^{8}\alpha^{2}+9216\delta^{2}\beta^{3}\alpha
\gamma-13824\delta\beta^{5}\alpha \gamma
-1728\beta^{7}\alpha\gamma-6144\delta\beta^{4}\gamma^{2}+1536\beta^{6}\gamma^{2}\]
\[-12288\delta\beta^{4}\alpha \epsilon+9216\beta^{6}\alpha
\epsilon+12288\beta^{5}\gamma
\epsilon-12288\delta\beta^{5}\zeta+18432\beta^{7}\zeta)N^{2}+(-384\delta^{4}\mu\beta^{2}\]
\[+1024\nu\delta^{3}\beta^{3}-3072\xi\delta^{2}\beta^{4}+256\delta^{3}\mu\beta^{4}-512\nu\delta^{2}\beta^{5}+3072\xi\delta
\beta^{6}+208\delta^{2}\mu\beta^{6}\]
\[-960\nu\delta \beta^{7}+960\xi\beta^{8}+288\delta
\mu\beta^{8}-192\nu\beta^{9}+\frac{129}{2}\mu\beta^{10}12288\beta^{6}K+3072\delta^{3}\beta^{2}\alpha^{2}-960\delta
\beta^{6}\alpha^{2}-\]
\[288\beta^{8}\alpha^{2}-9216\delta^{2}\beta^{3}\alpha
\gamma+960\beta^{7}\alpha \gamma
+6144\delta\beta^{4}\gamma^{2}+1536\beta^{6}\gamma^{2}12288\delta\beta^{4}\alpha
\epsilon \]
\[+6144\beta^{6}\alpha \epsilon-12288\beta^{5}\gamma
\epsilon+12288\delta\beta^{5}\zeta-6144\beta^{7}\zeta)N+(96\delta^{4}\mu\beta^{2}-256\nu\delta^{3}\beta^{3}+768\xi\delta^{2}\beta^{4}\]
\[+32\delta^{3}\mu\beta^{4}-64\nu\delta^{2}\beta^{5}-384\xi\delta\beta^{6}+20\delta^{2}\mu\beta^{6}+144\nu\delta\beta^{7}-144\xi\beta^{8}-54\delta
\mu\beta^{8}+36\nu\beta^{9}\]
\[-\frac{117}{8}\mu\beta^{10}-3072\beta^{6}K+768\delta^{4}\alpha^{2}-768\delta^{3}\beta^{2}\alpha^{2}-480\delta^{2}\beta^{4}\alpha^{2}+144\delta\beta^{6}\alpha+87\beta^{8}\alpha^{2}\]
\[-3072\delta^{3}\beta \alpha \gamma+2304\delta^{2}\beta^{3}\alpha
\gamma+960\delta\beta^{5}\alpha \gamma-144\beta^{7}\alpha
\gamma+3072\delta^{2}\beta^{2}\gamma^{2}\]
\[-1536\delta
\beta^{4}\gamma^{2}-576\beta^{6}\gamma^{2}+6144\delta^{2}\beta^{2}\alpha\epsilon-3072\delta
\beta^{4}\alpha\epsilon-2688\beta^{6}\alpha\]
\[\epsilon-12288\delta\beta^{3}\gamma
\epsilon+3072\beta^{5}\gamma\epsilon+12288\beta^{4}\epsilon^{2}-3072\delta\beta^{5}\zeta+768\beta^{7}\zeta\]
\section{\textbf{References}}
1. V.Fock, Z.Phys. 98, 145-154 (1935).
\newline
2. V.Bargmann, Z.Phys. 99, 576-582 (1936).
\newline
3. J.M.Jauch and E.L.Hill, Phys.Rev. 57, 641-645 (1940).
\newline
4. M.Moshinsky and Yu.F.Smirnov, The Harmonic Oscillator In Modern
Physics, (Harwood, Amsterdam, 1966).
\newline
5. J.Fris, V.Mandrosov, Ya.A.Smorodinsky, M.Uhlir and P.Winternitz, Phys.Lett. 16, 354-356 (1965).
\newline
6. P.Winternitz, Ya.A.Smorodinsky, M.Uhlir and I.Fris, Yad.Fiz. 4,
625-635 (1966). (English translation in Sov. J.Nucl.Phys. 4,
444-450 (1967)).
\newline
7. J.Drach, C.R.Acad.Sci.III, 200, 22 (1935), 200, 599 (1935).
\newline
8. S.Gravel and P.Winternitz, J.Math.Phys. 43(12), 5902 (2002).
\newline
9. S.Gravel, J.Math.Phys. 45(3), 1003-1019 (2004).
\newline
10. I.Marquette and P.Winternitz, J.Math.Phys. 48(1) 012902
(2007).
\newline
11. E.L.Ince, Ordynary Differential Equations, Dover, New York (1956).
\newline
12. Ya. I Granovskii, A.S. Zhedanov  and I.M. Lutzenko , J.
Phys. A24, 3887-3894 (1991).
\newline
13. P.Letourneau and L.Vinet, Ann.Phys. 243, 144 (1995).
\newline
14. D.Bonatsos, C.Daskaloyannis and K.Kokkotas, Phys.Rev. A48,
R3407-R3410 (1993).
\newline
15. D.Bonatsos, C.Daskaloyannis and K.Kokkotas, Phys. Rev.A 50, 3700-3709 (1994).
\newline
16. P.Letourneau and L.Vinet, Ann.Phys. 243, 144 (1995).
\newline
17. C.Daskaloyannis, J.Math.Phys. 42, 110 (2001).
\newline
18. C.Daskaloyannis, Generalized deformed oscillator and nonlinear
algebras, J.Phys.A: Math.Gen 24, L789-L794 (1991).
\newline
19. E.G.Kalnins, W.Miller and S.Post, SIGMA 4, 008 (2008). 
\newline
20. V.Sunilkumar, arXiv:math-ph/0203047 (2002).
\newline
21. V.Sunilkumar, B.A.Bambah and R.Jagannathan, J.Phys. A:Math.Gen.34, 8583 (2001).
\newline
22. V.Sunilkumar, B.A.Bambah and R.Jagannathan, mod. Phys. Lett. A17, 1559 (2002).
\newline
23. Y.A.Golfand and E.P.Likhtman, JETP Lett. 13, 323 (1971), A.Neveu and J.H.Schwarz, Nucl.Phys.B,31, 86 (1971), J.Wess and B.Zumino, Nucl.Phys.B70, 39 (1974), M.F.Sohnius,Phys.Rep.128,39 (1985).
\newline
24.  E.Witten, Nucl.Phys. B185, S13 (1981).
\newline
25. G.Darboux, C.R.Acad.Sci. Paris, 94, 1459 (1882)
\newline
26. T.F.Moutard, C.R.Acad.Sci. Paris, 80, 729 (1875), J.de L'ï¿½cole Politech., 45, 1 (1879).
\newline
27. E.Schrodinger, Proc.Roy. Irish Acad., 46A, 9 (1940), 47A, 53 (1941).
\newline
28. L.Infeld and T.E.Hull, Rev.Mod.Phys., 23, 21 (1951).
\newline
29. G.Junker, Supersymmetric Methods in Quantum and Statistical Physics, Springer, New York, (1995).
\newline
30. A.Khare and R.K.Bhaduri, Am.J.Phys, 62, 1008-1014 (1994)
\newline
31. L.Gendenshtein, JETP Lett., 38, 356 (1983).
\newline
32. C.Quesne, SIGMA 3, 067 (2007).
\newline
33. S.Gravel, ArXiv:math-ph/0310004 (2004).
\newline
34. B.Mielnik, J. Math. Phys. 25, 12  (1984).
\newline
35. Liu Ke-Jia, Chin. Phys. Soc,10, 4 (2001).
\newline
36. D.J.Fernandez C,V.Hussin and L.M.Nieto, J.Phys.A:Math. Gen. 27, 3547 (1994).
\newline
37. A.Jevicki and J.P.Rodrigues, Phys.Lett. B 146, 55 (1984).
\newline
38. J.Casahorran and J.G.Esteve, J.Phys.A:Math. Gen. 25 L347 (1992).
\newline
39. I.F.Marquez, J.Negro and L.M.Nieto, J.Phys. A:Math. Gen. 31, 4115 (1998).
\newline
40. A.Das and S.A.Pernice, Nucl.Phys.B 561, 357 (1999).
\newline
41. M.Znojil,Nucl.Phys.B662,554 (2003), M.Znojil, Phys.Lett. A 259, 220 (1999).
\newline
42. C.Bender, S.Boettcher and P.Meisinger, J.Math.Phys. 40, 2201 (1999).
\newline
43. A.Mostafazadeh, J.Math.Phys., 43, 205 (2002).
\newline
44. A.Sinha and P.Roy, Czech J.Phys. 54, 1, 129 (2004).
\newline
45. B.Bagchi, C.Quesne and M.Znojil, Mod.Phys.Lett. A 16, 2047 (2001).
\newline
46. P.E.Verrier and N.W.Evans, J.Math.Phys. 49, 022902 (2008).
\newline
47. M.A.Rodriguez, P.Tempesta and P.Wintertnitz, Phys. Rev. E78, 046608 (2008).
\newline
48. M.S.Plyushchay, Annals Phys. (N.Y.) 245, 339 (1996), M.Plyushchay, Int.J.Mod.Phys. A15, 3679 (2000), F.Correa and M.Plyushchay, Annals Phys.322, 2493 (2007).
\newline
49. F.Correa, L-M.Nieto and M.S.Plyushchay, Phys.Lett. B644, 94 (2007), F.Correa, V.Jakubsky, L-M.Nieto and M.S.Plyushchay, Phys.Rev.Lett. 101, 030403 (2008).
\newline
50. J.Hietarinta, Phys.Lett.A, 246, 1, 97 (1998).

\end{flushleft}
\end{document}